\documentclass[unnumsec,webpdf,contemporary,large]{oup-authoring-template}%

\graphicspath{{Fig/}}

\theoremstyle{thmstyleone}%
\theoremstyle{thmstyletwo}%
\theoremstyle{thmstylethree}%

\begin{document}


\journaltitle{Journal Title Here}
\DOI{DOI added during production}
\copyrightyear{YEAR}
\pubyear{YEAR}
\vol{XX}
\issue{x}
\access{Published: Date added during production}
\appnotes{Paper}

\firstpage{1}


\title[Meniscope]{Meniscope: A Low-Cost Fluid Interface Visualizer}

\author[1,$\ast$]{Daniel M. Harris\ORCID{0000-0003-2615-9178}}

\address[1]{\orgdiv{School of Engineering}, \orgname{Brown University}, \orgaddress{\street{Providence}, \state{RI}, \postcode{02912},  \country{USA}}}

\corresp[$\ast$]{Corresponding author. \href{email:daniel_harris3@brown.edu}{daniel\_harris3@brown.edu}}

\received{Date}{0}{Year}
\revised{Date}{0}{Year}
\accepted{Date}{0}{Year}



\abstract{In this work, we describe the development and application of a low-cost fluid interface visualizer referred to as the ``Meniscope.''  The device works using a color-based surface gradient detector method that maps the gradient of an air-water interface to a specific color on a target pattern below using a converging lens.  Sample experiments are outlined that showcase the working principle and functional versatility of the device.  The device and assembly instructions were piloted in a hands-on workshop, with pertinent feedback reviewed herein.  The Meniscope is a low-cost device that is capable of producing striking visualizations of static and dynamic free-surface deformations while introducing users to free-surface measurement techniques in an accessible and hands-on manner.
} 






\maketitle


\section{Introduction}

Capillary-scale interfacial deformations may regularly go unperceived by us but play an important role in a number of small-scale natural systems.  For instance, certain insects reside at the air-water interface and leverage the force of surface tension to balance their weight and remain afloat (\cite{vella2015floating}).  Their localized menisci interact with one another and their surroundings, and can be used to self-assemble (\cite{voise2011capillary,loudet2011mosquito,ko2022small}) or actively climb to shore (\cite{hu2005meniscus}).
Dynamic interactions with the interface generate propagating ripples known as capillary waves, which have been demonstrated to contribute in part to the propulsion of both animate (\cite{bush2006walking,buhler2007impulsive,roh2019honeybees}) and inanimate surface dwellers (\cite{couder2005walking,rhee2022surferbot,harris2025propulsion}). It has also been suggested that such waves can be used for communication and to locate prey at range by some species (\cite{bleckmann1985perception}), a topic that remains an active field of study (\cite{sbrocco2026lensing}).  In all of these examples the minuscule surface deformations are sub-millimetric, rendering careful non-intrusive measurement techniques critical to advancing our understanding of the natural world.

A number of non-intrusive techniques rely on the distinct optical properties of an immiscible fluid interface.  In particular, the difference in refractive index between air and water leads to a predictable distortion of light rays that pass through the interface.  Some methods that exploit the refractive properties of the free surface include shadowgraphy (e.g. \cite{roh2019honeybees}), free-surface synthetic Schlieren (\cite{moisy2009synthetic}), and color-based surface gradient detector methods (\cite{zhang1994measuring}).  While shadowgraphy has been popularized in various educational settings with wave tanks (e.g. PASCO Ripple Tank System), and the free-surface synthetic Schlieren has seen expanded application recently due to the release of open-source post-processing codes (\cite{wildeman2018real}) and additional technical improvements (e.g. \cite{metzmacher2022double}), the color-based surface gradient detector method has received relatively less attention.  While configuring, aligning, and calibrating the technique can be more cumbersome than other methods, the central qualitative effect can be realized with a few inexpensive components to create striking visualizations while introducing potential users to the method and some fundamental concepts in interfacial phenomena.

In this paper, we describe a low-cost device for visualizing the fluid interface in small-scale capillary-dominated systems, referred to as the ``Meniscope.''  First the fundamental optical principles underlying the working principle will be reviewed.  Following that, the design and fabrication steps of the device will be outlined, with sample introductory experiments detailed.  The successful use of the device in a recent hands-on workshop will also be described and future directions discussed.

\section{Fundamentals}
\begin{figure}[!t]%
\centering
\includegraphics{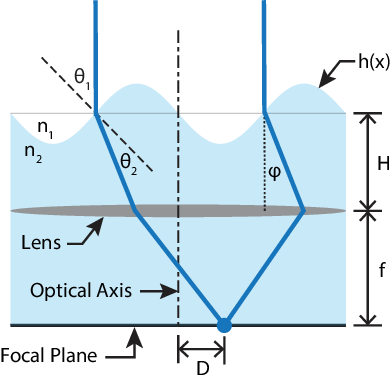}
\caption{Schematic highlighting the key optical properties underlying the surface gradient detector method (\cite{zhang1994measuring}).  A vertical ray first passes through a fluid interface with local slope $\frac{dh}{dx}$, thereby being reoriented by an angle $\varphi$. The ray then passes through a focusing lens with focal length $f$, focusing to a point on the focal plane a distance $D$ away from the central optical axis.  As such, an interface slope $\frac{dh}{dx}$ is mapped to a unique position $D$ on the focal plane.  In the absence of interfacial disturbances (i.e. $h(x)=H$), all incident rays map to the center of the focal plane, coincident with the optical axis of the lens.}\label{fundamentals}
\end{figure}

When light passes through an interface characterized by a change in the refractive index, its direction of propagation is changed.  The rule governing this deflection is known as Snell's law:
\begin{equation}
    \frac{\sin\theta_2}{\sin\theta_1}=\frac{n_1}{n_2}\label{eqn:snell}
\end{equation}
where $\theta_1$ is the angle of incidence (relative to the interface normal), $\theta_2$ is the outgoing angle, and $n_1$ and $n_2$ are the refractive indices of the interfacing media (Figure \ref{fundamentals}).  For the case of a ray passing from air (medium 1) to water (medium 2), $n_1=1$ and $n_2=1.33$, and $\theta_1\geq\theta_2$.  Note that the angle $\theta_1$ corresponds identically to the local angle of free surface (relative to the horizontal), and thus the magnitude and sign of $\theta_2$ uniquely encode the slope of interface via equation \ref{eqn:snell}.  In summary (and assuming small angles) a vertically incident ray is ultimately diverted from its initial path by an angle $\varphi$, which can be calculated as follows
\begin{equation}
    \varphi\approx\frac{dh}{dx}\left(1-\frac{n_1}{n_2}\right)\label{eqn:phi}
\end{equation}
where $\frac{dh}{dx}$ is the local slope of the free surface.  However to infer the slope in practice, we need a measurement of $\varphi$.  The color-based surface gradient detector method accomplishes this using a single converging lens and a 2D pattern placed at the focal distance ($f$) of the lens (Figure \ref{fundamentals}).  While most graphical depictions of a convex lens show normal incident rays focusing to a single point along the symmetry axis, rays that enter the lens at an oblique angle $\varphi$ also focus to the focal plane, but to a point a distance $D$ away from the central axis through the rule:
\begin{equation}
    D=f\tan\varphi.\label{eqn:D}
\end{equation}
Combining equations \ref{eqn:phi} and \ref{eqn:D} (and again assuming small angles) yields the final relation 
\begin{equation}
    \frac{dh}{dx}\approx{\frac{D}{f\left(1-\frac{n_1}{n_2}\right)}}.
\end{equation}
As such, if the distance $D$ can be measured (and $f$, $n_1$, and $n_2$ are known), one has direct access to the surface slope $\frac{dh}{dx}$.  The color-based surface gradient detector method accomplishes this by encoding each position $D$ with a unique color that can be visualized and recorded from above. The characteristic scale of the target pattern and the focal length of the lens conspire to determine the sensitivity of the method.

While this simple calculation was outlined for a 1D interface (i.e. $h(x)$), it is readily generalizable to 2D (i.e. $h(x,y)$), with the final result being a mapping between the local surface gradient ($\nabla h$) and the vector displacement on the focal plane relative to the central axis ($\bar{D}(x,y)$):
\begin{equation}
    \nabla h\approx{\frac{\bar{D}}{f\left(1-\frac{n_1}{n_2}\right)}}.\label{eqn:vec}
\end{equation}
Note that should any additional optical interfaces be present in the path, as is the case for the Meniscope, these must be accounted for as well and equation \ref{eqn:vec} appropriately revised.

\section{Device}

The Meniscope was designed to be a simple implementation of the color-based surface gradient detector method using low-cost and widely accessible materials.  The key components of the device include two plastic cups to set and maintain the working distances, a plastic Petri dish to hold the water, and three plastic credit-card-sized magnifying lenses.  The commercial magnifying lenses are Fresnel lenses, which consist of concentric rings of prisms to emulate the effect of a traditional convex lens but are much more compact and use far less material.  Three lenses were stacked to form a single optical unit in order to reduce the effective focal length $f$ and keep the device relatively compact.  Table \ref{tab1} summarizes these components with approximate costs.  A complete bill-of-materials, step-by-step construction guide, and sample experiments are provided in the freely available companion Instructables page associated with the device.\footnote{\url{https://www.instructables.com/Fluid-Interface-Visualizer/}} 

\begin{table}[!b]
\caption{Materials required for a single Meniscope, with current approximate costs in USD. Cost estimate assumes bulk purchase at typical available quantities at the time of writing.\label{tab1}}%
\begin{tabular*}{\columnwidth}{@{\extracolsep\fill}llll@{\extracolsep\fill}}
\toprule
Item & Number Required  & Cost \\
\midrule
Plastic Party Cup (16 oz.)    & 2   & \$0.26  \\
Credit-Card-Sized Magnifying Lenses    & 3   & \$0.57   \\
Plastic Petri Dish Base (90 mm diameter)    & 1   & \$0.36   \\
\midrule
{Total}    & \--   & \$1.19   \\
\botrule
\end{tabular*}
\end{table}

\begin{figure*}[!t]%
\centering
\includegraphics{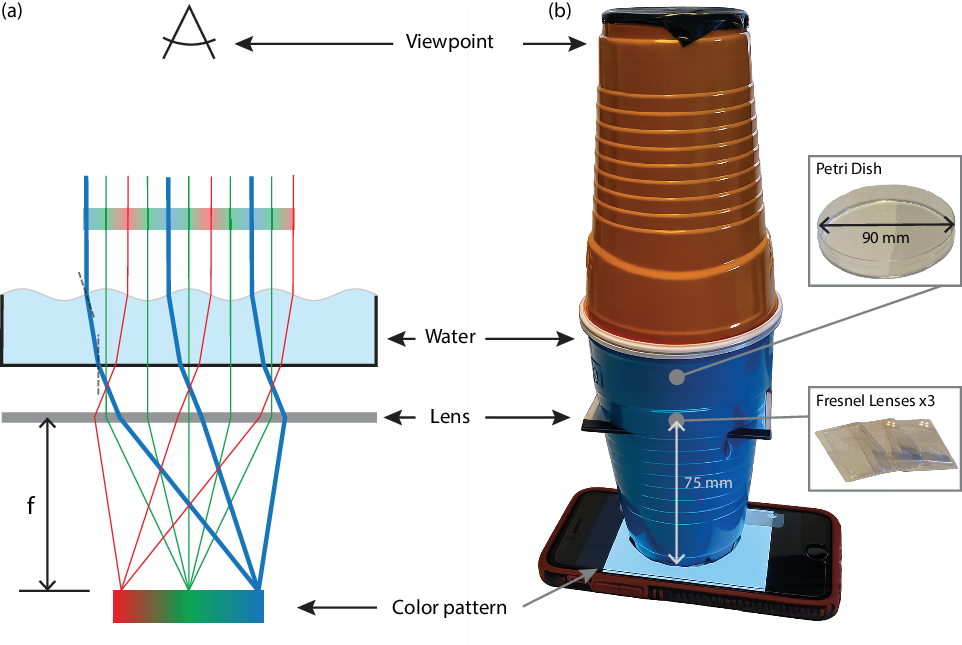}
\caption{(a) Schematic of the working principle behind the Meniscope, known as the color-based surface gradient detector method (\cite{zhang1994measuring}). When viewed from above, each surface gradient is mapped to a different point (encoded by color) on a screen placed at the focal plane of the lens. (b) Fully assembled device placed atop a cell phone displaying a target color pattern.}\label{schematic}
\end{figure*}

A schematic of the device as well as a labeled photo of an assembled unit are provided in Figure \ref{schematic}.  As can be seen, the lower cup holds both the water tray and the lens assembly, and rests atop the color pattern.  The upper cup shields the system from ambient light while also serving the dual purpose of setting an optimal viewing distance for the visualization. 

The only tool needed to fabricate the device is a cutting blade (e.g. razor or X-Acto knife).  With this tool, the ends of both cups are removed to provide optical access and diametrically opposed slots are cut into the side of one cup to hold the lenses in place.  Additionally, tape can be useful to hold some components together and cover sharp edges that might arise during cutting, but is optional.  Assuming a standard 16 oz. plastic party cup is used, the base plate of the Petri dish securely presses into the opening of the cup and does not require any adhesive to stay in place.  The target color pattern is loaded on a smartphone or tablet, which is assumed to be available to the user.  Alternatively an artist's light tracing pad can be used with color patterns printed on transparency paper.  

Once assembled, the next step is alignment.  In particular, when viewed directly from above at the working distance set by the upper cup, the lens stack should appear to be projecting a nearly uniform color across its surface.  When aligned appropriately, this color should be the color at the center of the target pattern.  If this is not the case, the device can be manually repositioned (or the image on the screen moved) to achieve better alignment.  Due to the practical limitations of the device and lenses, some aberrations will likely be present on the edges of the view.  

\subsection{Sample Experiments}

Three sample experiments were performed on the device that are easily reproducible and showcase different capillary-dominated fluid phenomena.

\begin{figure*}[!t]%
\centering
\includegraphics{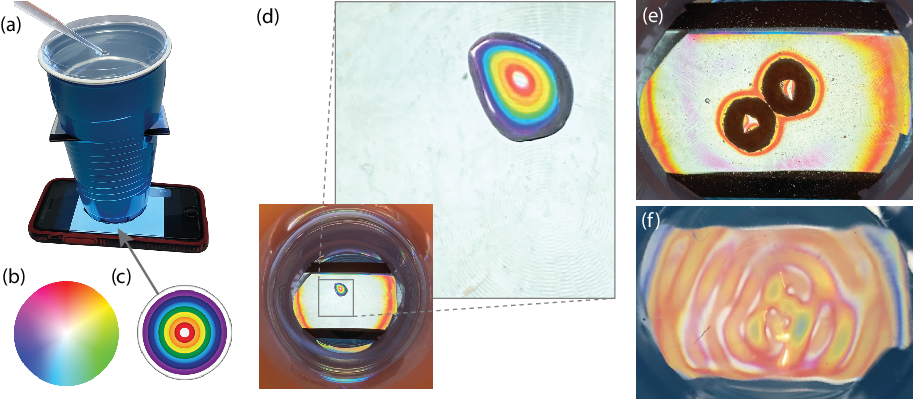}
\caption{(a) A small water droplet is deposited on the Meniscope Petri dish.  (b) Typical target color pattern used for the visualization technique that maps hue and saturation to gradient direction and magnitude, respectively. (c) Discrete color ``bulls-eye'' pattern used for example visualizations shown herein that was found to be easier to work with for demonstrative and exploratory purposes. (d) Image through the viewing cup of the sessile droplet. Away from the droplet, the color on the lens appears uniformly white, representing successful alignment with the center of the target pattern shown in (c). Some aberrations are visible near the edge of the lens due to practical limitations of the inexpensive optical setup. (e) Two Cheerios floating at an air-water interface are surrounded by a meniscus and attract one another through capillary attraction. (f) Periodic excitations lead to capillary waves forming at the air-water interface. White stripes correspond to regions of nearly zero slope, indicative of peaks and troughs of the excited wavefield.}\label{examples}
\end{figure*}

The first experiment is a sessile droplet, which is perhaps the easiest to perform.  A small droplet of water is deposited on the Petri dish as shown in Figure \ref{examples}(a).  While a more typical color pattern for 2D gradient mapping is shown in Figure \ref{examples}(b), we found that the discrete concentric ring pattern shown in Figure \ref{examples}(c) was easier to work with when getting started, and produced more immediately striking visual results.  Figure \ref{examples}(d) shows an example of a single sessile droplet visualized from above.  The background appears nearly uniformly white (consistent with successful alignment with the target pattern in Figure \ref{examples}(c)), except for the region occupied by the droplet.  The droplet samples a range of gradients spanning from zero (flat) at the center to a maximum slope corresponding to the contact angle with the dish.  If using a smartphone or tablet, the size of the target pattern can be changed in real-time which allows one to directly visualize how this scale influences the sensitivity of the measurement.  Additional discrete droplets of different sizes can also be added to the dish, with the maximum gradient (color) seen to be set by the contact angle rather than the size of the droplet itself.  This sample experiment represents a fully static capillary surface.

The second experiment (Figure \ref{examples}(e)) provides a visualization of capillary attraction, sometimes referred to as the ``Cheerios effect'' (\cite{vella2005cheerios}).  A freely floating object, in this case an actual Cheerio, is surrounded by a meniscus highlighted in color by the device.  The largest slopes occur nearest the cereal piece, and decay outwards.  The characteristic length scale of the meniscus is defined by the so-called capillary length, which is approximately 2.7 mm for an air-water interface in standard gravity conditions (\cite{delens20253d}).  A second Cheerio creates a nearly identical meniscus, and the two floating objects attract one another spontaneously in order to minimize the overall potential energy of the system. The capillary length also sets the characteristic length scale of the interaction between the freely floating cereal.  This experiment represents a dynamic interfacial problem, but one where the interface shape is typically assumed to evolve in a quasi-static manner. 

The third demonstrative experiment (Figure \ref{examples}(f)) is that of capillary waves, a fully dynamic process where fluid inertia and surface tension compete.  In this case, the support structure of the system (e.g. kitchen table for the example shown) is struck impulsively or periodically vibrated, and capillary waves form spontaneously at the free surface. The visualization in Figure \ref{examples}(e) shows a still image of an instantaneous wave field generated by periodically vibrating the table with a massage gun at a fixed frequency.  Here the peaks and troughs of the wavefield are highlighted in white (zero slope), with the largest surface slopes (and corresponding wave amplitudes) occurring near the center of the dish.  The location of the vibration source on the table can be moved relative to the device to change the strength of the forcing and thereby the wave amplitude.  In addition, the dominant wavelength of the waves can be modulated with the excitation frequency.

\begin{table*}[b]
\caption{Results from voluntary exit survey sent to participants of the ``Flow Visualization and Measurement at the Air-Water Interface'' workshop at SICB 2026.  During the hands-on portion of the workshop, each participant assembled and used a Meniscope. Of the approximately 30 participants in the workshop, $N=11$ responses to the survey were received. \label{tab2}}
\tabcolsep=0pt
\begin{tabular*}{\textwidth}{@{\extracolsep{\fill}}lcccccc@{\extracolsep{\fill}}}
\toprule%
& \multicolumn{5}{@{}c@{}}{Responses ($N=11$)} &  \\
\cline{2-6}%
Survey Question & 1 (Poor)  & 2 & 3 & 4 & 5 (Excellent) & Mean \\
\midrule
Overall, how would you rate the workshop?  & 0 & 0 & 0 & 3 & 8 & 4.7/5.0\\
How clear were the assembly instructions?  & 0 & 0 & 0 & 3 & 8 & 4.7/5.0\\
How effective was the device in demonstrating the concepts discussed?
  & 0 & 0 & 0 & 2 & 9 & 4.8/5.0\\
\botrule
\end{tabular*}
\end{table*}

\subsection{Pilot Workshop}
The Meniscope was first debuted as part of a workshop entitled``Flow Visualization and Measurement at the Air-Water Interface'' at the 2026 Society for Integrative and Comparative Biology Annual Meeting (SICB 2026). The workshop was two hours long in total, with the first hour dedicated to a more slide-based lecture-style overview of various techniques in the field.  The second hour was set aside for each participant to construct and use a Meniscope, following the instructions provided on Instructables (accessed via a QR code) and additional synchronous verbal guidance.  At the start of the event, each participant was provided with an unassembled kit and access to a razor and electrical tape for fabrication.  

A voluntary exit survey was distributed by email to all registrants following the workshop.  Of the approximately 30 participants who attended the hands-on portion of the workshop, 11 responded to the survey, with 10 of those reporting having completed a fully functional device in the time provided. Some additional relevant feedback is summarized in Table \ref{tab2}, with responding participants generally finding the assembly instructions clear and that the device was effective in demonstrating the concepts discussed in the preceding lecture portion.  One participant commented that the hands-on workshop was ``great hands on arts and crafts'' and that they were ``so happy to have a workshop with an activity,'' while another noted that the most valuable insight or skill they gained during the workshop was being able to ``make a working tool with common cups and others.''

\section{Discussion}

In this work, we have described the development of a low-cost fluid interface visualizer based on the color-based surface gradient detector method.  In addition, we have reviewed the essential optical principles enabling the technique and provided a number of sample experiments that can be readily performed with the device.  Through successful deployment at a hands-on workshop at SICB 2026, the device (and corresponding instructions) were validated to be effective for demonstrative purposes.  

Although not implemented at the workshop, the companion Instructables page describes an adaptation of the device for the shadowgraphy technique.  In this configuration, the color pattern is replaced by a  point-like source of light (LED or smartphone flashlight), and the lenses are used to collimate the light before it passes through the fluid bath. The transmitted light is then projected on a small translucent screen mounted above the Petri dish on another segmented plastic cup.  This technique is primarily sensitive to the interface curvature rather than its gradient, with peaks in the wavefield focusing the collimated light on the screen and thus appearing bright, and troughs providing the opposite effect and appearing dark.

Although the present device is not intended to provide quantitative measurements, future work might include development of a higher fidelity standalone version that would naturally move in this direction.  In addition to further standardizing certain aspects of the assembly, the setup would benefit from a dedicated light source and color pattern and improved alignment system, perhaps taking inspiration from other successful low-cost imaging devices such as the Foldscope (\cite{cybulski2014foldscope}) or the OpenFlexure microscope (\cite{collins2020robotic}).  Furthermore, open-source post-processing code for the color-based technique and a practical calibration guide would be extremely valuable towards broadening the user base.  Nevertheless, the device described herein can be rapidly assembled using commonly available resources and provides a natural introduction to the technique. Despite its limitations, it is capable of generating striking visualizations of the air-water interface, communicating the beauty and elegance of interfacial fluid mechanics to both seasoned experts and the previously uninitiated alike.

\section{Conflicts of interest}
The author declares that there are no competing interests.

\vspace{-0.1 in}
\section{Funding}
This work is supported in part by the National Science Foundation (NSF CBET-2338320). Financial support for materials for the pilot workshop at SICB 2026 was provided by La Vision, The Company of Biologists, and The Society for Integrative and Comparative Biology.

\vspace{-0.1 in}
\section{Data availability}
No data were created or analyzed in this study. A step-by-step build and use guide for the device is available at \url{https://www.instructables.com/Fluid-Interface-Visualizer/}.

\vspace{-0.1 in}
\section{Author contributions statement}
D.M.H. designed the device, performed the experiments, and wrote the manuscript.

\vspace{-0.1 in}
\section{Acknowledgments}
The author gratefully acknowledges Eli Silver and Emma Harris for support with the original prototype and testing, Robert Hunt for useful conversations related to the visualization technique, and Margaret Byron and Chris Roh for support with organizing, designing, and running the pilot workshop.


\bibliographystyle{oup-abbrvnat}
\bibliography{reference}

\end{document}